# Study of acoustic emission due to vaporisation of superheated droplets at higher pressure

Rupa Sarkar[*], Prasanna Kumar Mondal and Barun Kumar Chatterjee

Department of Physics, Bose Institute, 93/1, A P C Road, Kolkata-700009, India

**Abstract**

The bubble nucleation in superheated liquid can be controlled by adjusting the ambient pressure and temperature. At higher pressure the threshold energy for bubble nucleation increases and we have observed that the amplitude of the acoustic emission during vaporisation of superheated droplet decreases with increase in pressure at any given temperature. Other acoustic parameters such as the primary harmonic frequency and the decay time constant of the acoustic signal also decrease with increase in pressure. It is independent of the type of superheated liquid. The decrease in signal amplitude limits the detection of bubble nucleation at higher pressure. This effect is explained by the microbubble growth dynamics in superheated liquid.

## 1. Introduction

Superheated liquids are known to vaporise when irradiated by energetic particles since the invention of bubble chamber [1]. For over three decades superheated liquids have been used for the preparation of superheated emulsion detectors where superheated micro-droplets are dispersed in a viscoelastic gel (superheated droplet detector, SDD [2]) or in a soft polymer matrix (bubble detector, BD [3]). These detectors are used for the detection of neutrons, charged particles, gamma-rays etc. [2-5]. The SDD and BD are threshold type detectors where by changing the operating temperature and/or pressure the threshold can be controlled [2].

It is well known that whenever an energetic particle passes through the superheated liquid it deposits energy along its path and if the energy is sufficient then it can trigger the bubble nucleation [2, 6]. The bubble nucleation and subsequent vaporization of superheated droplet generate an acoustic pulse which can be detected by transducers [7]. Though different techniques have been used for the detection of droplet vaporisation in superheated emulsions [7-10], acoustic detection of nucleation is still one of the important techniques used in this field [11-14]. Acoustic detection is important because of its ability to detect vaporization of a single droplet, which also could enable one to identify the nature of the particle triggering the bubble nucleation [12-14]. This technique can be used in discriminating bubble nucleation events due to different types of radiation [12-14], which already has proved it's applicability not only in studying the bubble nucleation in SDD but also in the study of bubble nucleation events in detectors used in dark matter search experiment [14-15].

It was observed that the bubble nucleation rate decreases during the experiment at higher pressure [16] and despite visual observation of bubble formation it goes undetected by the transducer. This loss of count results in the reduction of the detection efficiency of the SDD. On the other hand the detector becomes more stable at higher pressure due to the increase of threshold energy for bubble nucleation. The study of nucleation at higher pressure is important to understand how the change in ambient pressure affects the bubble nucleation process in superheated droplets [16-18].

---
[*] Corresponding author: R. Sarkar; tel.: +91 33 23502403; fax: +91 33 23506790.
E-mail addresses: sarkar_rupa2003@yahoo.com (R. Sarkar),
prasanna_ind_82@yahoo.com (P.K. Mondal)
barun_k_chatterjee@yahoo.com (B.K. Chatterjee).

In this paper we have studied the acoustic emission during bubble formation in three different SDDs by using a $^{137}$Cs gamma-ray source. The experiments have been carried out by varying pressure to observe the effect of applied pressure on different acoustic parameters of sound waves emitted during bubble formation in superheated liquids. Here we have studied three different acoustic parameters of the sound wave viz. the amplitude, primary harmonic frequency and decay time constant. It is observed that with increase in ambient pressure at a given temperature all these parameters shifts towards lower values. It is also observed that the increase in ambient pressure reduces the number of acoustic pulses detected substantially. Here we discuss the probable reason of acoustic parameter shifts with the change in applied pressure on the basis of dynamics of microbubble growth in superheated liquids. The possible reason for the reduction of detected bubble nucleation events is also discussed here.

## 2. Theory

According to Seitz's thermal spike model [6], as the energetic radiation deposits energy along its path inside the superheated liquid, microbubbles are formed along the path. The bubble nucleation occurs if a microbubble reaches a critical size ($r_c$), and only then the vapour bubble can grow spontaneously to observable size. The energy needed to form a critical size microbubble is known as the threshold energy ($W$) for bubble nucleation which varies with temperature and pressure. For bubble nucleation to occur two conditions need to be satisfied: (i) the energy deposition must be greater than $W$, and (ii) this amount of energy must be deposited within a certain minimum distance inside the superheated liquid. The expression for $r_c$ and $W$ are given in Eq. (1) and Eq. (2) respectively [2, 6].

$$r_C = \frac{2\gamma}{(P_{SVP} - P)} \quad (1)$$

$$W = 2\pi r_C^2 (\gamma - T\frac{\partial \gamma}{\partial T}) - \frac{4}{3}\pi r_C^3 (P_{SVP} - P) + \frac{4}{3}\pi r_C^3 \rho_v h_v \quad (2)$$

Here $\gamma$ is the surface tension, $P_{SVP}$ is the saturation vapour pressure, $P$ is the ambient pressure, $T$ is the operating temperature, $\rho_v$ is the saturated vapour density, and $h_v$ is the latent heat of vaporization.

It is well known that the critical radius ($r_c$) and threshold energy ($W$) for bubble nucleation increase with increase in ambient pressure at a given temperature [16]. For superheated R-12 ($CCl_2F_2$, boiling point -29.8°C at atmospheric pressure) we have calculated the variation of $r_c$ with pressure at three different temperatures (Fig. 1). When a liquid is sufficiently superheated it can be used for the detection of low LET radiations like the gamma-rays. The gamma-rays deposit energy via their secondary electrons through Coulomb interactions.

The bubble nucleation and the subsequent vaporization of superheated droplet occur on a time scale of few microseconds [19]. The dynamics of bubble growth in superheated liquid and the acoustic emission is a complex phenomenon, which are subject of ongoing research [20]. After bubble nucleation the accelerating microbubble can produce a shockwave when its growth becomes supersonic in the metastable liquid. The newly formed bubble in SDD can oscillate with different harmonics which appeared to be triggered by the shockwave [21]. It is to be noted that in bubble chamber [1, 22], with a bulk volume of superheated liquid, the bubble nucleation and the subsequent vaporization of superheated liquid also produces acoustic emission [22]. Unlike SDD here the microbubble keeps on increasing and can vaporise the entire superheated liquid. Here there is no question of bubble oscillation and the acoustic emission is only due to the shock wave generated due to the supersonic growth of the microbubble. In SDD the rapid growth of the microbubble and the subsequent bubble oscillation produce a pressure pulse that spans for a few milliseconds. As has been mentioned earlier a piezoelectric transducer can be used for the detection of this acoustic signature of bubble formation.

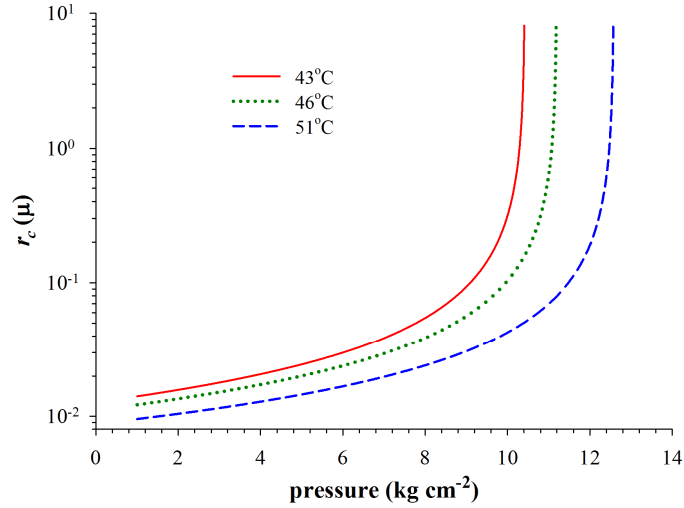

**Fig. 1.** For R-12 the variation of critical radius ($r_c$) for bubble nucleation with applied pressure at 43°C, 46°C, and 51°C.

## 3. Experiment

The experiments are performed using an in-house designed computer controlled high pressure manifold as shown in Fig. 2. The system is capable of pressurization up to about 43 kg cm$^{-2}$ with a precision of ±0.1 kg cm$^{-2}$. The pressure manifold consists of a storage tank $T_1$ which can be pressurized by a compressed air tank connected through a solenoid valve $S_1$, ball valves ($V_i$ and $V_1$) and a regulator (R). Here a needle valve is connected to the vent $V_t$, and a pressure gauge (PG) and a pressure sensor (WIKA R-1) reads the internal pressure. The WIKA R-1 pressure sensor is coupled with this system for automated pressure readout. A LabVIEW program was used for changing the pressure by controlling the solenoid valves $S_1$ and $S_2$. A round bottom thick walled glass vial containing the SDD is connected to the storage tank $T_1$ through the ball valve $V_2$ and solenoid valve $S_2$. The experiments are performed using three different types of SDDs consisting of micron sized droplets of R-12 ($CCl_2F_2$, boiling point -29.8°C at atmospheric pressure), R-134A ($C_2H_2F_4$, boiling point -26.3°C at atmospheric pressure) and R-1216 ($C_3F_6$, boiling point -29.4°C at atmospheric pressure).

The detectors are fabricated using the simple emulsification technique reported earlier in Mondal et al. [23]. In brief, first a viscoelastic gel is prepared by mixing the glycerol with commercial ultrasound gel in a suitable proportion such that it can hold the droplets in suspension. We have also added a surfactant Tween 80 to the gel (0.1% of the gel by volume) to improve the detector stability. The gel is then degassed for few hours to remove the air pockets. About 200 ml of this degassed gel is taken in a pressure tight container, where a measured amount of low boiling point liquid is also transferred under pressure. The amount of liquid taken varies from liquid to liquid. After that the gel and liquid is sheared with the help of an electric stirrer, which breaks the liquid into small droplets. After shearing the droplets are brought to the superheated state by slowly releasing the container pressure. The emulsion is then poured into glass vials and stored in refrigerator. The droplet size distribution can be controlled to some extent by controlling the stirring time and stirring speed. The detectors used in this work are prepared with a stirring speed of 1400 rpm for about 10 minutes. The droplet size distributions of different SDDs are measured using an optical microscope. For R-12, R-134A and R-1216 SDDs the average droplet size are respectively 12.2±4.2 μm, 11.3±4.4 μm and 7.9±2.3 μm.

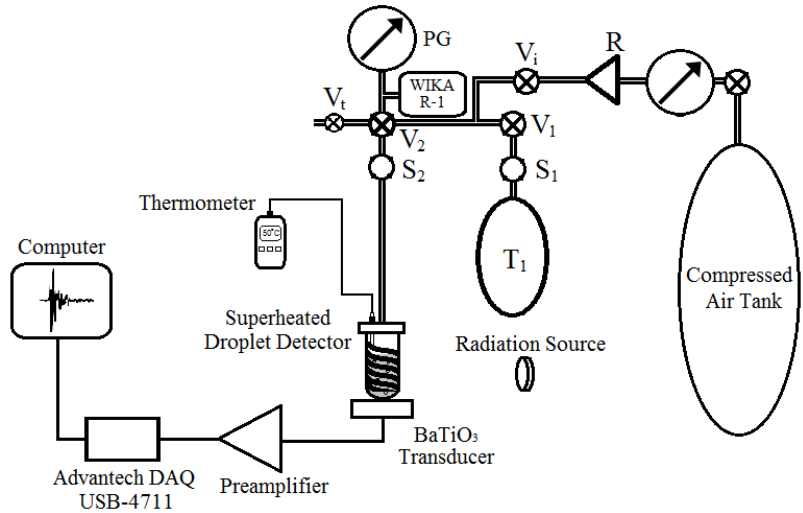

**Fig. 2.** Schematic representation of the experimental set up using a pressure manifold and data acquisition system.

To study the effect of applied pressure on different parameters of the acoustic pulses emitted during vaporization of superheated droplets, the experiments are performed at different pressure keeping the temperature constant. The temperature of the detector is controlled by a heating coil connected to a programmable power supply and a temperature controller system. At a fixed temperature the pressure is varied from 1.0 kg cm$^{-2}$ to a higher pressure where the detector was still gamma-ray sensitive. The pressure is varied in steps of 0.7 kg cm$^{-2}$. For R-12 detector at temperatures 43.2°C, 46.2°C and 51.1°C the maximum applied pressures are 1.7 kg cm$^{-2}$, 2.4 kg cm$^{-2}$ and 4.5 kg cm$^{-2}$ respectively. We have studied the R-134A detectors at 44.9°C, 47.8°C and 51.5°C, where the maximum applied pressure are 3.8 kg cm$^{-2}$, 4.5 kg cm$^{-2}$ and 5.2 kg cm$^{-2}$ respectively. Similarly for R-1216 detector at temperatures 28.7°C, 32.3°C and 36.1°C the experiments are performed at a maximum pressure of 2.4 kg cm$^{-2}$, 3.1 kg cm$^{-2}$, 4.5 kg cm$^{-2}$ respectively. To induce the bubble nucleation, the SDD is exposed to gamma-rays from $^{137}$Cs gamma-ray source.

The detector is first equilibrated at the desire temperature for about 20 minutes at an elevated pressure to avoid the unwanted droplet vaporization during temperature rise and then the detector is pressurized to a desired value. The applied pressure over the detector at any given temperature is adjusted such that the gamma-ray induced nucleation rate in that condition is substantial. As shown in Fig. 2, the detector vial is placed on a BaTiO$_3$ transducer (National Physical Laboratory, India, central frequency 220 kHz) which detects the acoustic signal generated during droplet vaporization. The transducer converts the acoustic signal into an electrical signal, which is then amplified and acquired in a computer with a sampling rate of 100 kS/s using an Advantech USB-4711 data acquisition card (DAQ) operating in Labview platform. The bandwidth of the amplifier is 125 kHz, with a maximum gain of about 10$^3$. To avoid electrical noise a threshold voltage of 150 mV is used as the trigger which rejects all the pulses with maximum amplitude below this threshold. The experiment is then repeated at a different pressure at the same temperature. Here at each pressure about 800 waveform data for droplet vaporization events are recorded. To avoid the experimental systematic error the pressure over the system was set in a random order. This eliminates the possible effect of the diffusion of superheated liquid into the gel which could change the droplet size distribution at high pressure.

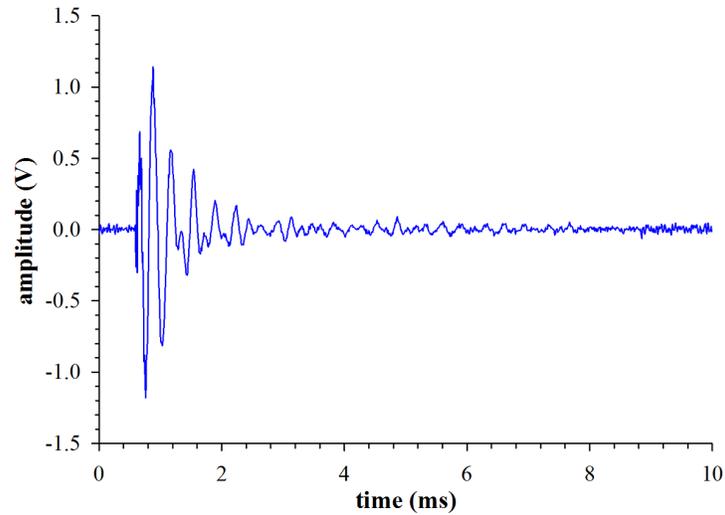

**Fig. 3.** A typical acoustic signal due to vaporization of superheated R-134A droplet at pressure 3.1 kg cm$^{-2}$ and temperature 51.5°C.

A typical waveform recorded during vaporization of superheated R-134A droplet at a pressure 3.1 kg cm$^{-2}$ and temperature 51.5°C is shown in Fig. 3. As shown in Fig. 3, at the beginning the signal waveform has a fast rise and then it falls slowly as damped oscillations. These waveform data are used for the measurement of different acoustic parameters.

**4. Observation**

In this work the waveform data are analyzed using LabVIEW program in order to explore different acoustic parameters.

**4.1 Variation of the decay time constant of the signal with operating pressure**

To obtain the decay time constant ($\tau$) of the signal, first the amplitude envelope of the waveform is calculated using the Hilbert transform of the waveform [11, 13]. The decaying part of the envelope is then fitted to an exponential function $y(t) = Ae^{-t/\tau}$, where $t$ is the time and A is a constant. Here the best fitted curve gives the decay time constant of the waveform. For all the pulses $\tau$ is obtained separately which shows a variation from detector to detector (Fig. 4). For a particular type of detector, at a given operating temperature and pressure, $\tau$ usually shows a distribution [13]. The average values of $\tau$ at different pressure for R-12, R-134A and R-1216 SDDs at temperatures 51.1°C, 51.5°C and 36.1°C are shown in Fig. 4. Here we observe that for all the SDDs at a given operating temperature $\tau$ initially decreases with increase in operating pressure up to about 3 kg cm$^{-2}$ and then it does not change much with increase in pressure. The decay time constant for all the three SDDs studied here at other temperatures show similar behavior as Fig. 4 (data not shown here).

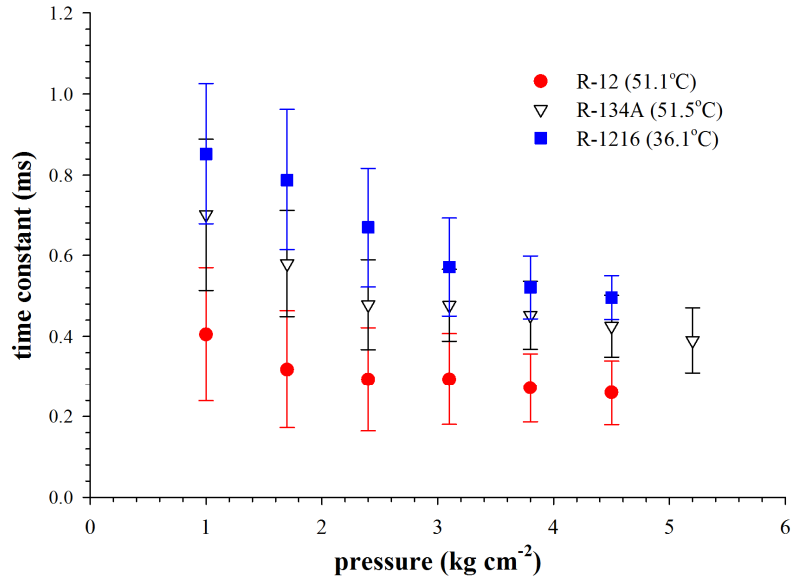

**Fig. 4.** The variation of decay time constant of the waveforms with operating pressure for R-12, R-134A and R-1216 at temperatures 51.1°C, 51.5°C and 36.1°C respectively.

**4.2 Variation of the primary harmonic of the signal with operating pressure**

It is well known that the acoustic signal from the SDDs has a well defined frequency. Here the primary harmonic frequency ($f$), defined as the frequency corresponding to the highest peak in the fast Fourier transform (FFT) of the waveform, is obtained for all the acquired waveform data. For a particular SDD at a given operating condition the primary harmonic shows a distribution [13]. Here we observe that pulses from different SDDs have primary harmonics in different frequency regions [Fig. 5]. The variation of average primary harmonic frequency with operating pressure for R-12, R-134A and R-1216 SDDs at temperatures 51.1°C, 51.5°C and 36.1°C are shown in Fig. 5. As evident from Fig. 5, at a given temperature the primary harmonic frequency decreases with increase in pressure up to about 3 kg cm$^{-2}$ and then it does not change much with increase in pressure. The primary harmonic frequency for all the three SDDs studied here at other temperatures show similar behavior as Fig. 5 (data not shown here).

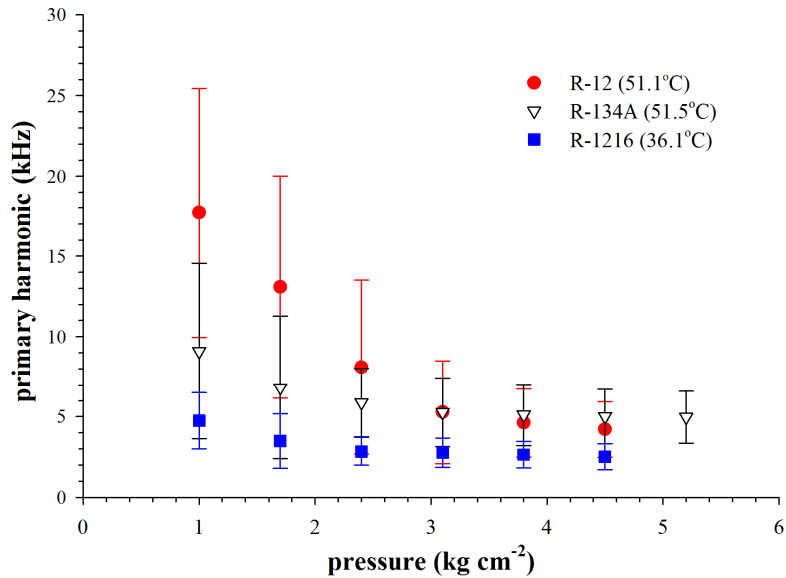

**Fig. 5.** The variation of primary harmonic frequency of the waveforms with operating pressure for R-12, R-134A and R-1216 at temperatures 51.1°C, 51.5°C and 36.1°C respectively.

**4.3 Variation of pulse amplitude of the signal with operating pressure**

To obtain the amplitude distribution for all the signals acquired in a set of experiment the maximum amplitude of each waveform is obtained using a LabVIEW program. The amplitude distribution of the pulses generated due to the vaporization of superheated R-1216 droplets at different pressure and at temperature 36.1°C is shown in Fig. 6. The amplitude distributions for R-1216 at 28.7°C and 32.3°C, and for R-12 and R-134A show similar behavior as Fig. 6 (data not shown here).

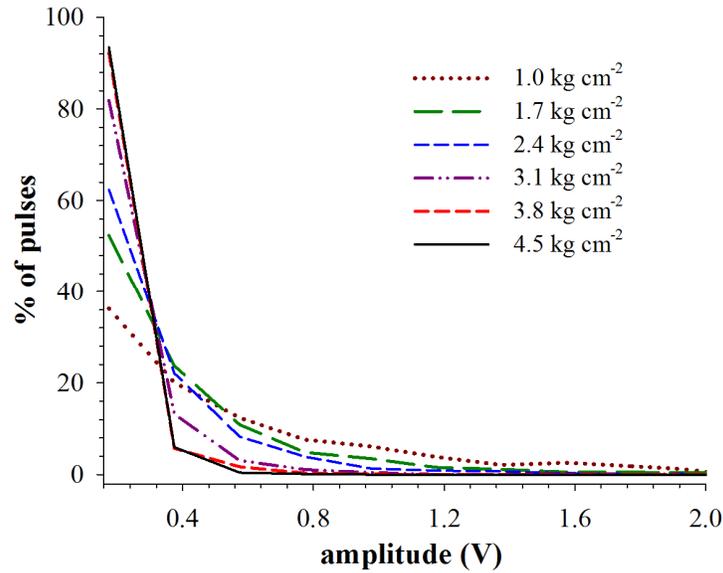

**Fig. 6.** The amplitude distribution of the pulses due to vaporization of superheated R-1216 droplets at different pressures at temperature 36.1°C.

As observed in Fig. 6, the amplitude distribution of acoustic signal moves towards the lower amplitude region with the increase in applied pressure. During experiments at higher pressure it is also observed that the accumulated number of pulses are much lower compared to the number of bubbles observed in the detector vial. It indicates that as we move to the higher pressure the percentage of lower amplitude pulses increases, i.e. the intensity of the sound energy emitted during bubble formation in superheated liquid decreases with increase in ambient pressure. To understand the effect of applied pressure on the amplitude of sound emitted during bubble formation in superheated liquid, the percentage of pulses greater than a reference voltage is calculated at each pressure at a given temperature and plotted as a function of pressure as shown in Fig. 7-9. This gives us an estimate of how many events will go undetected if we set a different reference voltage other than 150 mV for the detection of acoustic pulses. Here Fig. 7 represents the percentage of pulses with amplitude greater than the reference voltage of 250, 450, and 650 mV as a function of pressure for superheated R-1216 at 36.1°C. As evident from Fig. 7, for a detection threshold of 250 mV 36% of the pulse will go undetected at 1 kg cm$^{-2}$ because here only about 64% pulses have amplitudes higher than 250 mV. Here at 4.5 kg cm$^{-2}$ the number of pulses going undetected raises to about 94 % for the detection threshold 250 mV. Similarly for R-12 (51.1°C) and R-134A (51.5°C) the plots obtained with different reference voltage are shown in Fig. 8-9.

As evident from Fig. 7-9, the percentage of pulses with amplitude greater than a threshold value decreases monotonically with increase in pressure irrespective of the nature of the superheated liquid. At other temperatures too the data shows similar variations for all the three liquids investigated here.

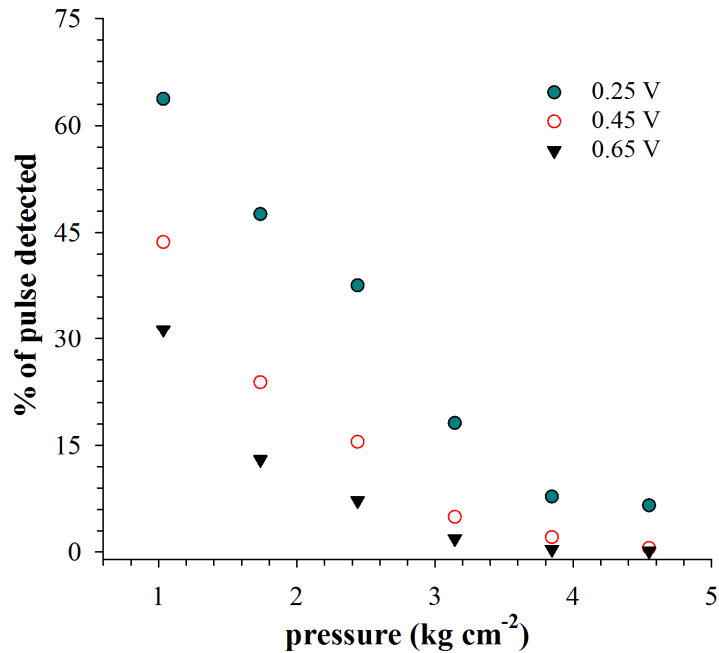

**Fig. 7.** For superheated R-1216 at 36.1°C the percentage of pulses with amplitude greater than a reference voltage at different pressure.

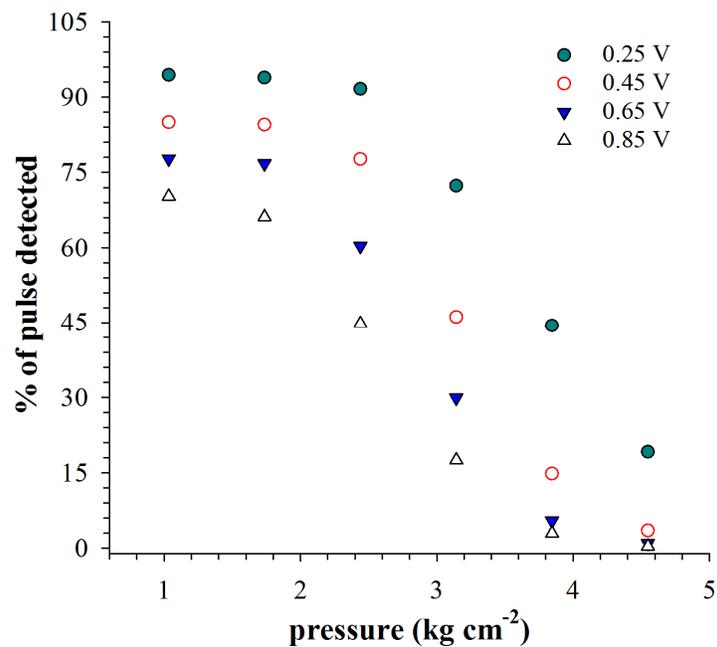

**Fig. 8.** For superheated R-12 at 51.1°C the percentage of pulses with amplitude greater than a reference voltage at different pressure.

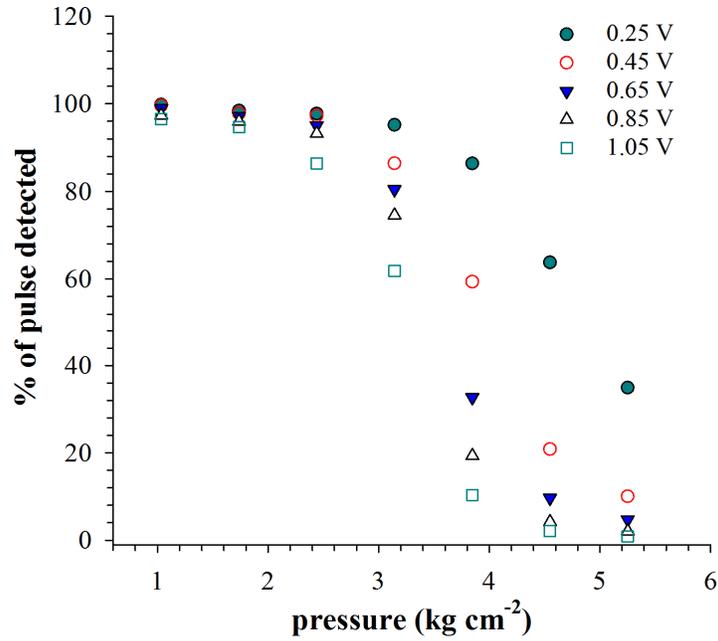

**Fig. 9.** For superheated R-134A at 51.5°C the percentage of pulses with amplitude greater than a reference voltage at different pressure.

**5. Discussions:**

We have studied the effect of increasing pressure on different acoustic parameters of the acoustic pulses emitted during vaporization of superheated droplets. Here we have observed that the decay time constant, primary harmonic frequency and amplitude of the acoustic pulses decreases with increase in operating pressure at a given temperature for all the three SDDs studied here. The decay constants are found to be different for different SDDs (Fig. 4) with a reduction of about 40% are observed within the pressure range studied here. The fact that the decay constant decreases with increase in operating pressure is in consistent with that reported in Felizardo et al. [11], where $C_3F_8$ SDDs were studied up to 2 bar.

Here we observe that the primary harmonic frequency (*f*) is also different for different SDDs. The R-1216 SDDs have the lowest *f* values compared to R-12 and R-134A SDDs. For R-12 SDDs there is a decrease of about 80% in *f* values compared to those of R-134A and R-1216 SDDs for which the change is about 45% in the pressure range studied here. In literature there are reports of both increase [11] and decrease [24] of *f* values with increase in applied pressure. Here for all the three SDDs we have observed the reduction of *f* values with increase in pressure. This is due to the excitation of lower bubble oscillation modes at higher pressure. In our earlier publication we have reported that in SDDs there can be excitation of different modes of bubble oscillation [21] induced by the shockwave generated during the expansion of microbubble in superheated droplet. During bubble expansion a shock wave can only be generated when the bubble expansion velocity exceeds the speed of sound in superheated liquid. The speed of sound in superheated liquid is usually much lower than that in normal liquid [25] and it increases with the increase in applied pressure [26-27] at a given temperature. Since with increase in pressure both the $r_c$ and *W* increase at a given temperature, here at higher pressure the radial velocity of the growing microbubble will cross the velocity of sound at a larger bubble radius compared to that at lower pressure. Larger the radius of the microbubble at which the radial velocity crosses the velocity of sound, smaller is the energy of the microbubble and hence smaller will be the energy of the shock wave. A shock wave with lower energy can only trigger bubble oscillation with lower order modes, as we observe here at higher pressure.

Here it is observed that the increase in ambient pressure reduces the amplitude of the acoustic emission during nucleation. This can also be explained from the dynamics of bubble nucleation in superheated liquid. As we move to the higher pressure $W$ increases and the growing microbubbles overcome this larger barrier at a greater critical radius ($r_c$) as shown in Fig. 1. Larger the $r_c$ smaller is the portion of the remaining liquid to be vaporised after bubble nucleation and hence lower is the stored energy which will be emitted as an acoustic emission due to the bubble oscillation. Here at higher pressure the energy of the shock wave and the stored energy are small compared to that at lower pressure. This leads to the emission of low amplitude acoustic signal at higher pressure.

Since with increase in pressure the acoustic signal amplitude decreases the detection threshold voltage plays an important role for the detection of acoustic pulses as evident from Fig. 7-9. Here we observe that as we increase the detection threshold more and more pulses go undetected at any given pressure. For R-1216 the fraction of pulses detected decrease monotonically with increase in pressure (Fig. 7), and for R-12 and R-134A we have a plateau region (Fig. 8-9) in the low pressure region. This is due to the fact that for R-1216 the average amplitude of the acoustic pulses is much smaller compared to R-12 and R-134A. Here the average pulse amplitude for R-1216, R-12 and R-134A at 1 kg cm$^{-2}$ pressure are 0.8 V (at 36.1$^o$C), 2.1 V (at 51.1$^o$C) and 4.4 V (at 51.5$^o$C) respectively. The relatively high pulse amplitude for R-12 and R-134A is due to the effect of elevated temperature and larger droplet size. In the plateau region the fraction of pulse detected does not change much with increase in applied pressure since most of the pulses have amplitudes greater than the threshold voltage.

As has been mentioned in the experiment section that we need to set a detection threshold voltage of 150 mV during the experiment to avoid the electrical noise received during experiments. For this reason there was no pulses recorded below 150 mV. Here as the operating pressure is increased the pulse height shifts towards lower values and more and more pulses go undetected. At some point the pressure would be so high that all the pulses will have amplitude lower than 150 mV and here no signal will be detected. Hence beyond this limiting pressure we will have virtually silent bubble nucleation, i.e. droplet vaporisation without any detectable acoustic emission, which would limit the detection of bubble nucleation at that pressure at a given temperature. To overcome this problem, detection system other than the acoustic detection need to be developed which will not limit the detection of bubble nucleation at elevated pressure.

**Acknowledgements**

R. Sarkar thanks the Science and Engineering Research Board (SERB, Government of India) for research grant (SR/FTP/PS-115/2013). The authors also express their gratitude to Dr. S. Gupta Bhattacharyya, Bose Institute, for the use of optical microscope for droplet size measurement.

**References:**
[1] D. A. Glaser, Phys. Rev. 87, (1952) 665.
[2] R. E. Apfel, Nucl. Instrum. Meth. 162, (1979) 603.
[3] H. Ing, H.C. Birnboim, Nucl. Tracks Radiat. Meas. 8 (1-4), (1984) 285.
[4] S.L. Guo, L. Li, B.L. Chen, T. Doke, J. Kikuchi, K. Terasawa, M. Komiyama, K. Hara, T. Fuse, Nucl. Instrum. Meth. B 198, (2002) 135.
[5] F. d'Errico, R. Nath, Med. Phys. 27 (2), (2000) 401.
[6] F. Seitz, Phys. Fluids 1, (1958) 2.
[7] R.E. Apfel, S.C. Roy, Rev. Sci. Instrum. 54, (1983) 1397.
[8] F. d'Errico, A.D. Fulvio, M. Maryañski, S. Selici, M. Torrigiani, Radiat. Meas. 43, (2008) 432.
[9] B. Roy, B.K. Chatterjee, S.C. Roy, Radiat. Meas. 29 (2), (1998) 173.

[10] P.K. Mondal, B.K. Chatterjee, Meas. Sci. Technol. 19, (2008) 105802.
[11] M. Felizardo, R.C. Martins, A.R. Ramos, T. Morlat, T.A. Girard, F. Giuliani, J.G. Marques, Nucl. Instrum. Meth. A 589, (2008) 72.
[12] A.D. Fulvio, J. Huang, L. Staib, F. d'Errico, Nucl. Instrum. Meth. A 784, (2015) 156.
[13] P.K. Mondal, S. Seth, M. Das, P. Bhattacharjee, Nucl. Instrum. Meth. A 729, (2013) 182.
[14] F. Aubin, et al., New J. Phys. 10, (2008) 103017.
[15] M. Felizardo, et al., Phys. Rev. Lett. 105, (2010) 211301.
[16] M. Das, T. Sawamura, Nucl. Instrum. Meth. A 536, (2005) 123.
[17] R. Sarkar, M. Datta, B.K. Chatterjee, Nucl. Instrum. Meth. A 682, (2012) 31.
[18] P. Rezaeian, G. Raisali, A. Akhavan, Radiat. Meas. 83, (2015) 31.
[19] J.E. Shepherd, B. Sturtevant, J. Fluid Mech. 121, (1982) 379.
[20] A.J. Robinson, R.L. Judd, Int. J. Heat Mass Trans. 47, (2004) 5101.
[21] P.K. Mondal, B.K. Chatterjee, Phys. Lett. A 375, (2011) 237.
[22] D.V. Jordana, et al., Appl. Radiat. Isot. 63, (2005) 645.
[23] P.K. Mondal, R. Sarkar, B.K. Chatterjee, Appl. Radiat. Isot. 90, (2014) 1.
[24] M. Felizardo, et al., Astro. Phys. 49, (2013) 28.
[25] R.A. Aziz, C.C. Lim, D.H. Bowman, Can. J. Chem. 45, (1967) 1037.
[26] B.A. Younglove, J. Res. Natl. Bur. Stand. 86 (2), (1981) 165.
[27] P. Kiełczyński, M. Szalewski, S. Piekarski, arXiv:1306.1012, (2013).